\documentclass[fleqn,twoside]{article}
\usepackage{amsfonts}
\usepackage{latexsym}
\usepackage{gc,cite}
\heads
{K.A. Bronnikov and S.V. Grinyok}
{Electrically charged and neutral wormhole instability in scalar-tensor
gravity}

\def\mn{_{\mu\nu}}
\def\MN{^{\mu\nu}}

\def\og{{\overline g}}

\def\M{{\mathbb{M}}}

\def\R{{\mathbb R}}
\def\S{{\mathbb S}}

\def\Str{\mbox{$\S_{\rm trans}$}}
\def\ME{\mbox{$\M_{\rm E}$}}
\def\MJ{\mbox{$\M_{\rm J}$}}

\def\umx{u_{\rm max}}
\def\phio{\phi_0}

\def\GR{general relativity}
\def\sph{spherically symmetric}
\def\ssph{static, spherically symmetric}

\def\wh{wormhole}
\def\whs{wormholes}

\bls{1.02}
\begin{document}
\twocolumn[
\prepno{gr-qc/0509062}{\GC {11} 75--81 (2005)}

\vspace{1cm}

\Title {Electrically charged and neutral wormhole instability\yy
        in scalar-tensor gravity}

\Authors
{K.A. Bronnikov\foom 1} {and S.V. Grinyok\foom 2}
{Centre for Gravitation and Fundam. Metrology, VNIIMS,
        3-1 M. Ulyanovoy St., Moscow 119313, Russia;\\
 Institute of Gravitation and Cosmology,
        PFUR, 6 Miklukho-Maklaya St., Moscow 117198, Russia}
{Bauman Moscow State Technical University, Moscow, Russia}


\Abstract
{We study the stability of static, spherically symmetric, traversable
wormholes with or without an electric charge, existing due to conformal
continuations in a class of scalar-tensor theories with zero scalar field
potential (so that Penney's or Fisher's well-known solutions hold in the
Einstein conformal frame). Specific examples of such wormholes are those
with nonminimally (e.g., conformally) coupled scalar fields. All boundary
conditions for scalar and metric perturbations are taken into account. All
such wormholes with zero or small electric charge are shown to be unstable
under spherically symmetric perturbations. The instability is proved
analytically with the aid of the theory of self-adjoint operators in Hilbert
space and is confirmed by numerical computations.}


] 
\email 1 {kb20@yandex.ru}
\email 2 {stepanv1@mail.ru}

\section{Introduction}

   Lorentzian wormholes as hypothetic macroscopic or astrophysical objects
   are of great interest from the viewpoint of possible causality violation
   (time machines etc.) \cite{Morris-Thorne, MTY} and from an observational
   viewpoint, as specific scatterers of stellar, galactic and quasar
   radiation \cite{Shats}. From the viewpoint of gravitation theories, they
   are striking examples of extremely strong gravitational fields free of
   singularities.

   A search for traversable \wh\ solutions to the gravitational field
   equations with realistic matter has been for long, and is still remaining
   to be, one of the most intriguing challenges in gravitational studies.
   One of attractive features of \whs\ is their ability to support electric
   or magnetic ``charge without charge'' \cite{wheeler} by letting the lines
   of force thread from one spatial asymptotic to another.

   As is widely known, traversable wormholes can only exist with exotic
   matter sources, more precisely, if the energy-momentum tensor (EMT) of
   the matter source of gravity violates the local and averaged null energy
   condition (NEC) $T\mn k^{\mu}k^{\nu} \geq 0$, $k_\mu k^\mu =0$
   \cite{hoh-vis}. It is known, for instance, that nonlinear electrodynamics
   with any Lagrangian of the form $L({\cal F})$, ${\cal F} = F\MN F\mn$,
   coupled to \GR, cannot produce a \ssph\ \wh\ metric \cite{br01-ned}.
   Though, an effective \wh\ geometry for electromagnetic wave propagation
   can appear as a result of the electromagnetic field nonlinearity
   \cite{nov1, nov2}.

   Scalar fields are able to provide good examples of matter needed for
   \whs: on the one hand, in many particular models they do exhibit exotic
   properties, on the other, many exact solutions are known for gravity with
   scalar sources. We will consider some examples of charged \wh\ solutions
   in the presence of massless scalar fields.

   Let us begin with the action for a general (Bergmann-Wagoner) class of
   scalar-tensor theories (STT), where gravity is characterized by the
   metric $g\mn$ and the scalar field $\phi$ in the presence of the
   electromagnetic field $F\mn$ as the only matter source:
\bearr            \nq\,
    S = \int d^4 x \sqrt{g}\{ f(\phi) R [g]             \label{act}
                 + h(\phi)g\MN\phi_{,\mu}\phi_{,\nu} - F\MN F\mn\}.
\nnn
\ear
   Here $R[g]$ is the scalar curvature, $g = |\det (g\mn)|$, $f$ and $h$
   are certain functions of $\phi$, varying from theory to theory. Exact
   \ssph\ solutions for this system are well known \cite{penney, br73}, but
   their qualitative behaviour is rather diverse and depends on the nature
   of the functions $f$ and $h$.

   Wormholes form one of the generic classes of solutions in theories where
   the kinetic term in (\ref{act}) is negative \cite{br73} (more precisely,
   if $l(\phi)$, defined in (\ref{ps-f}), is negative). A particular case
   of this kind of wormholes, namely, wormholes with a ``ghost''
   massless minimally coupled scalar field in \GR\ [\eq(\ref{act}),
   $f(\phi)\equiv 1,\ h(\phi)\equiv -1$] was considered by H. Ellis
   \cite{h_ellis}.

   The energy conditions, NEC in particular, are, however, violated as well
   by ``less exotic'' sources, such as the so-called nonminimally coupled
   scalar fields in \GR, represented by the action (\ref{act}) with the
   functions
\beq                   				         \label{nonmin}
      f(\phi) = 1-\xi \phi^2, \quad \xi =\const; \qquad h(\phi) \equiv 1.
\eeq

   Scalar-vacuum (with $F\mn=0$) \ssph\ \wh\ solutions were found in such a
   theory in \Ref{br73} (and were recently discussed in \Ref{bar-vis99})
   for conformal coupling, $\xi=1/6$, and in \Ref{bar-vis00} for any $\xi >
   0$. The easiness of violating the energy conditions, so evident due to
   the appearance of \wh\ solutions, even made Barcel\'o and Visser discuss
   a ``restricted domain of application of the energy conditions''
   \cite{bar-vis00}. This class of \whs\ exists in theories with scalar
   fields possessing a normal kinetic term but which admit a conformal
   continuation \cite{vac4}. The latter means that a singularity in the
   Einstein frame maps to a regular sphere in the Jordan frame, and the
   latter may be smoothly continued beyond this sphere. In \wh\ solutions,
   the second spatial asymptotic occurs in this new region of the Jordan
   manifold.

   We have recently proved \cite{worm3} that all these scalar-vacuum
   wormhole solutions are unstable under \sph\ perturbations:  it turns out
   that there exists at least a single mode of growing physically meaningful
   perturbations. The characteristic time of their growth is of the order of
   the time needed for a photon to cover a length equal to the \wh\ throat
   radius \cite{worm3}. (Our earlier results \cite{worm1,worm2}, according
   to which the perturbation growth rate is unlimited in the linear
   approximation, were obtained without taking into account the smoothness
   requirement for metric perturbations and therefore need revision.)

   The purpose of this paper is to extend these results to charged \whs.
   We first discuss the background configurations, namely, charged \ssph\
   \whs\ which appear in scalar-tensor theories (STT) of gravity
   in which the effective gravitational constant can change its sign due to
   conformal continuation \cite{vac4}. The investigation is, however,
   restricted to massless fields for which Penney's well-known solution
   \cite{penney} (or Fisher's \cite{fish} in the case of zero charge) holds
   in the Einstein frame. As examples, we describe scalar-electrovacuum \wh\
   solutions of the theory (\ref{act}), (\ref{nonmin}).

   Then we examine the stability problem, including the behaviour of metric
   perturbations related to those of the scalar field. A physically
   meaningful metric perturbation of an initially regular configuration
   should be regular everywhere. This requirement turns out to impose an
   additional constraint on the scalar field perturbations, which makes the
   stability problem quite nontrivial. We prove that, for sufficiently small
   electric charges, there exists at least a single growing mode of
   physically meaningful perturbations, i.e., such \whs\ are unstable, with
   roughly the same increment of perturbation growth as for similar
   neutral \whs.

   Finally, we briefly report the results of numerical studies. They confirm
   the instability conclusion and show that, as the charge grows, the
   perturbation increment decreases, indicating \wh\ stabilization for
   larger charges.

\section{Charged wormhole solutions}

\subsection {The general static solution}

   The general STT action (\ref{act}) is
   simplified by the well-known conformal mapping \cite{Wagoner1970}
\beq
   g\mn = \og\mn/|f(\phi)|,                             \label{trans-g}
\eeq
   accompanied by the scalar field transformation $\phi\mapsto \psi$ such
   that
\beq                                                       \label{ps-f}
   \frac{d\psi}{d\phi}= \pm \frac{\sqrt{|l(\phi)|}}{f(\phi)},
      \qquad    l(\phi) \eqdef fh +\frac 32
                        \biggl(\frac{df}{d\phi}\biggr)^2.
\eeq
   In terms of $\og\mn$ and $\psi$ the action takes the form
\bearr                                                   \label{act-E}
    S = \int d^4 x \sqrt{|\og|} \Bigl\{ (\sign f) \Bigl[  R [\og]
\nnn\cm
    + \og \MN \psi_{,\mu} \psi_{\nu} \,\sign l(\phi)\Bigr]
                        - F^{\mu\nu}F_{\mu\nu}   \Bigr\}
\ear
   (up to a boundary term which does not affect the field equations). Here
   $R[\og]$ is the Ricci scalar obtained from $\og\mn$, and the indices are
   raised and lowered using $\og\mn$. The electromagnetic field Lagrangian
   is conformally invariant, and $F\mn$ is not transformed.

   The space-time $\MJ[g]$ with the metric $g\mn$ is referred to as the
   {\it Jordan conformal frame}, generally regarded to be the physical
   frame in STT; the {\it Einstein conformal frame\/} $\ME[\og]$ with the
   field $\psi$ then plays an auxiliary role. The action (\ref{act-E})
   corresponds to conventional \GR\ if $f>0$, and the normal sign of scalar
   kinetic energy is obtained for $l(\phi) > 0$.

   The general \ssph\ solution to the Einstein-Maxwell-scalar equations that
   follow from (\ref{act-E}), was first found by Penney \cite{penney} and in
   a more complete form in \cite{zay, brpr}. Let us write it in the form
   suggested in \cite{br73}, restricting ourselves to the ``normal'' case
   $f >0$, $l>0$:
\bear
    ds_{\rm E}^2  \eql \e^{2\gamma(u)} dt^2
        -\e^{2\alpha(u)}du^2 - \e^{2\beta(u)} d\Omega^2       \label{ds-E}
\nnn \nqq
   = \frac {q^{-2}dt^2}{s^2(h, u+u_1)}
   - \frac{q^2 s^2(h,u+u_1)}{s^2(k,u)}
    \biggl[\frac{du^2}{s^2(k,u)}+d\Omega^2\biggr],
\nnn \\
     \psi(u)\eql Cu + \psi_1,                              \label{psi}
\\
    F_{01} \eql - F_{10} =
       q \,\e^{\alpha + \gamma - 2\beta}                    
\nn
      \eql \frac{1}{q\, s^2(h, u+u_1)},                     \label{F01}
\ear
   where the subscript ``E'' stands for the Einstein frame;
   $d\Omega^2= d\theta^2 + \sin^2\theta d\varphi^2$ is the linear element
   on a unit sphere; $q=q_e$ (the electric charge), $C$ (the scalar charge),
   $h$, $k$ and $\psi_1$ are real integration constants.  The function
   $s(k,u)$ is defined as follows:
\beq                                                         \label{s}
    s(k,u) = \vars     {
                    k^{-1}\sinh ku,  \ & k > 0 \\
                                 u,  \ & k = 0 \\
                    k^{-1}\sin ku,   \ & k < 0.  }
\eeq

   Here $u$ is a convenient radial variable (it is a harmonic coordinate
   in the Einstein frame, $\DAL u =0$). The range of $u$ is $0 < u < \umx$,
   where $u=0$ corresponds to spatial infinity, while $\umx$ may be finite
   or infinite depending on the constants $k$, $h$ and $u_1$.

   The integration constants are related by
\bear                                                       \label{khC}
    2k^2 \sign k \eql 2h^2 \sign h + C^2,
\\
    s^2(h,\ u_1) \eql 1/q^2.                                  \label{u1}
\ear
   The latter condition, preserving some discrete arbitrariness of $u_1$,
   provides the natural choice of the time scale ($\og_{00}=1$) at spatial
   infinity ($u=0$). Without loss of generality we put $C > 0$ and $\psi_1=0$.

   As usual, in addition to the electric field $F_{01}=-F_{10}$
   given by (\ref{F01}), one can include a radial magnetic field
   $F_{32}=-F_{23} = q_m \sin\theta$ where $q_m$ is the magnetic charge.
   One should then understand $q^2$ in (\ref{ds-E}), (\ref{u1}) and
   further on as $q^2 = q_e^2 + q_m^2$; in (\ref{F01}) one should replace
   $q$ with $q_e$ in the first line and $1/q$ with $q_e/q^2$ in the second
   line. In what follows, we will bear in mind this opportunity without
   special mentioning.

   The solution contains four essential integration constants:
   $k$ or $h$ and the charges $q_e,\ q_m$ and $C$. The mass $M$
   in the Einstein frame is obtained by comparing the asymptotic of
   (\ref{ds-E}) at small $u$ with the Schwarzschild metric:
\beq
     GM = \pm \sqrt{q^2 + h^2 \sign h}                         \label{GM}
\eeq
   where $G$ is Newton's gravitational constant. The ``$\pm$'' sign
   depends on the choice of $u_1$ among the variants admitted by (\ref{u1}).

   The Reissner-Nordstrom solution of \GR\ is a special case obtained
   herefrom by putting $C=0$. Then from (\ref{khC}) it follows $h=k$, and the
   familiar form of the Reissner-Nordstrom metric is recovered after
   a transition to the curvature coordinates,
   $-\og_{\theta\theta} = r^2$:
\beq
    r = \frac{|q|\,s(k,u+u_1)}{s(k,u)} \then
    \e^{2ku} = \frac{r+k-GM}{r-k-GM}.                    \label{RN}
\eeq

   To obtain another special case $q=0$ (the scalar-vacuum solution), one
   should consider the limit $q\to 0$ preserving the boundary condition
   (\ref{u1}). This is only possible for $k > h \geq 0$ and $u_1\to\infty$.
   The resulting metric is
\beq
   ds_{\rm E}^2 = \e^{-2hu} dt^2 - \frac{k^2 \e^{2hu}}{\sinh^2 (ku)}
      \biggr[\frac{du^2}{\sinh^2(ku)} + d\Omega^2 \biggl].   \label{g-vac}
\eeq
   The scalar field is determined, as before, from (\ref{psi}), and the
   integration constants are related by
\beq                                                      \label{k-vac}
            2k^2 = 2h^2 + C^2
\eeq
   It should be noted that in (\ref{g-vac}), (\ref{k-vac}) the constant $h$
   can have any sign, and for the mass $M$ we have simply $GM = h$.

   This is the Fisher solution \cite{fish} in terms of the harmonic $u$
   coordinate. Its more familiar form, used, in particular, in
   Refs.\,\cite{bar-vis99, bar-vis00}, corresponds to the coordinate $r$
   connected with $u$ by $r = 2k/(1-\e^{-2ku})$, and
   the metric in terms of $r$ has the form
\bearr
             ds^2_{\rm E}= (1-2k/r)^a dt^2              \label{g-vac'}
    \nnn \quad
         - (1-2k/r)^{-a}\bigl[dr^2 + r^2(1-2k/r)d\Omega^2\bigr],
\ear
   with $a=h/k$. The Schwarzschild solution is then recovered in case $C=0$,
   $a=1$.

   All the corresponding Jordan-frame solutions for $l(\phi)>0$ are obtained
   from (\ref{psi}), (\ref{ds-E}) using the transformation (\ref{trans-g}),
   (\ref{ps-f}).

\subsection {Continued solution in the Jordan frame}

   Let us now turn to wormhole solutions for the nonminimal coupling
   (\ref{nonmin}), $\xi > 0$. The transformation (\ref{ps-f}) takes the form
\beq                                                      \label{trans-f}
   \frac{d\psi}{d\phi}
            = \frac {\sqrt{|1-\phi^2(\xi-6\xi^2)|}}{1-\xi\phi^2},
\eeq
   where, without loss of generality, we have chosen the plus sign before
   the square root. We assume that spatial infinity in the Jordan space-time
   $\MJ$ corresponds to $|\phi| < 1/\sqrt{\xi}$, where $f(\phi) >0$, so that
   the gravitational coupling has its normal sign.

   Generically, the solution in $\ME[\og]$ has a naked singularity at
   $u=\umx$, and, though its nature can change due to the transformation to
   $g\mn$, it remains to be a singularity in $\ME[g]$. An exception is the
   case when the solution is smoothly continued in $\MJ[g]$ through the
   sphere \Str\  ($u=\infty$, $\phi=1/\sqrt{\xi}$) which is singular in
   $\ME[\og]$ but regular in $\MJ[g]$. The infinity of the conformal factor
   $1/f$ thus compensates the zero of both $\og_{tt}$ and
   $\og_{\theta\theta}$ simultaneously. Wormhole solutions can only be
   found in this case. It happens when, in accord with (\ref{khC}),
\beq                                                        \label{k2h}
    k = 2h =2C/\sqrt{6} >0,\cm    u_1>0,
\eeq
   which selects a special subfamily among all solutions. We will restrict
   our attention to this subfamily. Note that now
   $s(k,u)= (2h)^{-1}\sinh (2hu)$, \
   $s(h, u+u_1) = h^{-1}\sinh (hu+hu_1)$ and $\umx=\infty$. According to
   (\ref{psi}) and (\ref{trans-f}),  we have $\psi\to\infty$ as $\phi\to
   1/\sqrt{\xi}-0$.

   Under the condition (\ref{k2h}) the solutions with and without charge in
   \ME\  are conveniently written in isotropic coordinates. Indeed, putting
   $y = \tanh (hu)$, we obtain:
\bear             \nq                   \label{dsEy}
     ds_{\rm E}^2 \eql \frac{(1-y^2) y_1^2}{(y+y_1)^2} \biggl[ dt^2
        -h^2\frac{(y+y_1)^4}{y_1^4\, y^4}(dy^2 + y^2 d\Omega^2)\biggr],
\nnn
\\
    \psi \eql \frac{\sqrt{6}}{2} \ln \frac{1+y}{1-y},      \label{psi-y}
\\                                                          \label{F-y}
    F_{01} \eql - F_{10} = \frac{q_e}{h}\,\frac{y_1^2}{(y+y_1)^2},
\ear
   where
\beq
    y_1 = \tanh (hu_1) = \frac{h}{\sqrt{h^2 + q^2}}.       \label{y1}
\eeq
   The vacuum solution is included here as the special case $q=0$, $y_1 =1$.
   The range of $u$, $u\in \R_+$, is converted into $y\in (0,1)$ where $y=0$
   corresponds to spatial infinity and $y=1-0$ to a naked singularity.

   To proceed to the Jordan frame, let us integrate
   \eq(\ref{trans-f}). This gives \cite{bar-vis00}%
\footnote
        {We have changed the notations as compared with
        \cite{bar-vis00}, in particular, we have replaced
        $\Phi_{\xi} \mapsto \sqrt{6}\phi$, $H\mapsto 1/H$ and $F^2\mapsto
    1/B$, to avoid imaginary $F$ at $\phi>1/\sqrt{\xi}$.}
\beq                                                          \label{ln}
        \psi = - \sqrt{3/2}\ln [B(\phi)H^2 (\phi)]
\eeq
    where
\beq                                                            \label{B}
   B(\phi) = B_0
        \frac{\sqrt{1-\eta\phi^2} - \sqrt{6}\xi\phi}
             {\sqrt{1-\eta\phi^2} + \sqrt{6}\xi\phi},
\eeq
    $B_0=\const$, while $H(\phi)$ is different for different $\xi$:
\bearr \nq                                                       \label{H}
     0<\xi<1/6:
\nnn
     H(\phi) = \exp\left[-\frac{\sqrt{1-6\xi}}{\sqrt{6\xi}}
                \arcsin \sqrt{\eta}\phi\right],
\nnn  \nq
     \xi > 1/6:
\nnn
     H(\phi) = \left[\sqrt{-\eta}\,\phi
                    +\sqrt{1-\eta\phi^2}\right]
                    ^{\fract{\sqrt{6\xi-1}}{\sqrt{6\xi}}},
\ear
    where $\eta=\xi(1-6\xi)$, and $H\equiv 1$ for $\xi=1/6$. The function
    $H(\phi)$ is finite in the whole range of $\phi$ under consideration.

    \eq (\ref{ln}) is valid for $\phi<1/\sqrt{\xi}$, and the Jordan-frame
    metric $g\mn = \og\mn/f$ under the condition (\ref{k2h}) can be written
    in terms of the coordinate $y$ as follows:
\bearr                                                      \label{ds-J}
     ds^2_{\rm J} =
               \frac{BH^{2}}{1-\xi\phi^2}
                \biggl[ \frac{(1+y)^2}{(y+y_1)^2}y_1^2dt^2
\nnn \cm
        -h^2\frac{(1+y)^2 (y+y_1)^2}{y_1^2\, y^4}
                (dy^2 + y^2 d\Omega^2)\biggr],
\ear
    where $y$ can be expressed in terms of $\phi$:
\beq
    y = \frac{1-BH^2}{1+BH^2}.                              \label{yBH}
\eeq

    The metric is thus actually expressed in terms of the scalar field
    $\phi$ used as a coordinate. The isotropic coordinate $y$ conveniently
    shortens the expression (\ref{ds-J}) and makes it easy to see that the
    metric, originally built for $\phi < \phio$ ($y<1$), is smoothly
    continued across the surface \Str\ ($\phi=\phio,\ y=1$). Indeed, in a
    close neighbourhood of \Str, for $\phi = (\phi - \delta)/\sqrt{\xi}$
    with $\delta \ll 1$ one has
\[
     B \approx B_0 \delta/(12\xi),\qquad
     1 -\xi\phi^2 \approx 2\delta\,
\]
    whence
\beq                                                         \label{factor}
     \frac{BH^2}{1-\xi\phi^2}\ \bigg|_{y=1} =
            \frac{B_0}{24\xi} H^2\bigg|_{\phi=\phio}.
\eeq
    It is easily shown that this ratio is not only finite on \Str\ but also
    smoothly changes across it, so that \eq (\ref{ds-J}) comprises an
    analytic continuation of the metric, obtained from
    (\ref{ds-E})--(\ref{F01}) in case (\ref{k2h}) by the transformation
    (\ref{trans-g}), (\ref{ps-f}), beyond \Str. The coordinate $y$ covers
    the whole manifold $\MJ[g]$, and it is now possible to study the
    properties of the system as a whole.

    Before doing that, let us note that the new region $\phi > \phio$
    ($y>1$) in \MJ\ can also be obtained by the same transformation
    (\ref{trans-g}), (\ref{ps-f}) from a certain Einstein frame.  An
    essential difference from the previous solution is that, since
    $f(\phi)$ is now negative, (\ref{act-E}) leads to the Einstein equations
    with a reversed sign of the electromagnetic energy-momentum tensor. As a
    result, the solution in this second Einstein-frame manifold%
\footnote
       {The prime will designate quantities describing the Einstein frame
         or $\phi > 1/\sqrt{\xi}$.  }
    $\ME'$ will have the same form (\ref{ds-E})--(\ref{F01}), but with the
    replacement
\beq
    s(h,u+u_1)\ \mapsto \ h'{}^{-1} \cosh (h'u + h'u_1),    \label{E'1}
\eeq
    where $h'>0$, and the relation (\ref{khC}) is replaced by $2k'{}^2 =
    2h'{}^2 + C'{}^2$ where $k' >0$.

    The solution in $\ME'$ is also regularized by the factor $1/f$ on
    \Str, and the integration constants in it satisfy the condition
    $k'=2h'$, similar to (\ref{k2h}). Other integration constants are
    adjusted as well, in particular, the charges $q_e$ and $q_m$ are the
    same on both sides of \Str, providing the continuity of the
    electromagnetic field.

\subsection{Wormhole solutions}

    Let us begin with the simplest case $\xi=1/6$ (conformal coupling).
    Then instead of (\ref{ln})--(\ref{H}) one can write for $\phi< \sqrt{6}$
\beq
    \phi = \sqrt{6}\tanh [(\psi + \psi_0)/\sqrt{6}],
            \cm \psi_0 = \const,               \label{phi-1/6}
\eeq
    where $\psi=Cu$ and due to (\ref{k2h}) $C= h\sqrt{6}$.
    The Jordan-frame solution in terms of the isotropic
    coordinate $y$ takes the form \cite{br73}
\bear
     ds_{\rm J}^2 \eql \frac{(1+yy_0)^2}{1-y_0^2}           \label{ds6}
      \biggl [\frac{y_1^2\, dt^2}{(y+y_1)^2}
\nnn \cm\
    -h^2\,\frac{(y+y_1)^2}{y_1^2 y^4}(dy^2 + y^2 d\Omega^2)\biggr],
\\
     \phi \eql \sqrt{6}\, \frac{y+y_0}{1 + yy_0},           \label{phi6}
\ear
    where $y_0 = \tanh (\psi_0/\sqrt{6})$ and $y_1 \in (0,1)$;
    the expressions for $F\mn$ are evident.

    The original Einstein-frame solution corresponds to $y < 1$, $y=0$ is
    spatial infinity while the sphere $y=1$ is \Str, where the solution
    (\ref{ds6}), (\ref{phi6}) is manifestly regular. The region $y>1$ is an
    analytic continuation of the solution in $\MJ[g]$ to $\phi > \sqrt{6}$
    and corresponds to another Einstein-frame solution described above.

    The properties of the solution at $y > 1$ depend on the constant $y_0$
    which characterizes the $\phi$ field at spatial infinity. Namely, if
    $y_0 < 0$, then the solution has a naked singularity at $y = -1/y_0 >1$.
    If $y_0=0$, we obtain a black hole with electromagnetic and scalar
    charges \cite{bbm70, brpr, br73, bek74}; introducing $r= h(y+y_1)/(y_1
    y)$, we obtain
\bear
     ds^2 \eql (1-m/r)^2 dt^2
     			- (1-m/r)^{-2}dr^2 -r^2 d\Omega^2,   \label{BH}
\nn
     \phi \eql C/(r-m)
\ear
    where $m= GM = \sqrt{h^2 + q^2},\ C = \sqrt{6}h$.
    On the horizon, $r=m$, despite $\phi\to\infty$, the energy-momentum
    tensor of the scalar field is finite. This solution (mainly its
    neutral special case $q=0$) was repeatedly discussed as an interesting
    counterexample of the well-known no-hair theorems; its instability
    under \sph\ perturbations has been proved in Ref.\,\cite{br78}.

    Lastly, if $y_0 >0$, then $y$ ranges from 0 to
    $\infty$, and $y=\infty$ is another flat spatial infinity. This is the
    sought-for \wh\ solution, parametrized by the four constants $h$, $q_e$,
    $q_m$ and $y_0$. The position and radius of the \wh\ neck (minimum of
    $r^2 = -g_{\theta\theta}$) are given by
\beq                                                            \label{neck}
      y_{\rm neck} = \frac{\sqrt{y_1}}{\sqrt{y_0}}, \cm
      r_{\rm neck} =
         \frac{h (1 + \sqrt{y_0 y_1})^2}{y_1\sqrt{1-y_0^2}} .
\eeq

    For $\xi \ne 1/6$ the analytical relations are much more complicated,
    but the qualitative behaviour of the solution can be described
    rather easily.

    In case $\xi > 1/6$, for any $B_0$, with growing $\phi$ the quantity
    $B^2 H^{-4}$ eventually reaches the value 1, where $g_{\theta\theta}\to
    \infty$, i.e., we arrive at another spatial asymptotic, and it is
    straightforward to verify that this infinity is flat. In other words, we
    obtain again a static \wh.

    In case $\xi < 1/6$ everything depends on $B_0$. If
\beq                                                           \label{B0}
    B_0 < B_0^{\rm cr}
           =\exp\biggl(-\pi\sqrt{\frac{1-6\xi}{6\xi}}\biggr),
\eeq
    the situation is the same as for $\xi> 1/6$, i.e., a \wh. If
    $B_0 > B_0^{\rm cr}$, then, while $g_{\theta\theta}$ is still finite,
    $\phi$ reaches the value $1/\sqrt{\eta} = 1/\sqrt{\xi(1-6\xi)}$, the
    location of a curvature singularity \cite{bar-vis00}. So we have a naked
    singularity instead of a \wh. Lastly, for $B_0 = B_0^{\rm cr}$, the
    maximum value of $\phi$ is again $1/\sqrt{\eta}$, but now it is
    non-flat spatial infinity.

\section{Stability analysis}

\subsection{Problem setting}

\def\Srd{Schr\"odinger}

    The present stability analysis repeats the main features of the similar
    analysis for electrically neutral \whs\ \cite{worm3}. We will use the
    results obtained there. In particular, the \Srd\ form of the equations
    is defined for functions of the same Hilbert space as for uncharged
    \whs. Meanwhile, the boundary value problem now depends on two
    parameters: the energy-like parameter (actually, the increment of
    perturbation growth) and the \wh\ charge. The \Srd\ operator, if written
    in a more or less visually graspable form, turns out to be
    non-self-adjoint, which considerably reduces our ability to study its
    properties. We shall solve the problem for small charges only, with the
    aid of perturbation theory for operators' point spectra. At small
    charges our operator evidently turns into its analogue for uncharged
    \whs.

    Let us write down the linearized Einstein equations for perturbations
    $\delta\alpha$, $\delta\beta$, $\delta\gamma$ (as in \cite{worm3}, we
    are working in the Einstein picture), taking the metric (\ref{dsEy}) as
    the background one:
\bearr                 		                     \label{E_pertQ}
    \e^{2\gamma}R_1^0 = 2(\delta\dot{\beta}'
                	- \beta'(\delta\dot\beta + \delta\dot\gamma) -
                        	\gamma'\delta\dot\beta)=0,
\nnn
    \e^{2\alpha}R_2^2 = 2\beta''(2\delta\beta + \delta\gamma)
\nnn
    - 2\e^{2\beta+2\gamma}\delta\beta + \e^{4\beta}\delta \ddot \beta
                        - \delta\beta'' = -4 q^2 \e^{2\gamma}\delta\beta.
\ear
    The primes here denote $\d/\d y$. The scalar field perturbation is absent
    since we have used the gauge $\delta\phi \equiv 0$. This gauge is
    manifestly physical (i.e., the perturbations do not comprise a pure
    gauge) and is particularly convenient for describing connections between
    the Jordan and Einstein pictures \cite{worm3}.

    As in \cite{worm3}, we carry out variable separation with the
    exponential $\e^{\Omega t}$ and arrive at a boundary-value problem with
    containing the separation constant $\Omega$ as an eigenvalue. The set of
    perturbation equations reduces to a single equation for
    $\delta\beta(y,t)$ expressing the dynamics of the only existing degree
    of freedom:
\beq                                                  \label{st1q}
    \delta\beta'' - \Omega^2\ \delta\beta\ s^4(y) + F(y)\ \delta\beta'
                                  + G(y)\ \delta\beta=0 ,
\eeq
    where $s$, $F$, $G$ are functions of $y$ obtained from the metric
    (\ref{dsEy}):
\bear
       F(y) \eql -2\beta''/\beta',
\nn
       G(y) \eql -2\beta''+ 2\beta''\gamma'/\beta' + 2\e^{2\beta+2\gamma}
        -4 q^2 \e^{2\gamma}\delta\beta ,
\nn
       s(y) \eql \e^{\beta}.
\earn
    The boundary conditions will be discussed a little later.

    A few words about choosing $y$ as a radial coordinate and using it in a
    whole range covering two different Einstein pictures before and beyond
    \Str\ ($y=1$). The point is that if one considered the perturbation
    problem in each Einstein frame separately (using, e.g., the harmonic
    coordinate $u$), it would be necessary to study two different operators
    and prove that their spectra coincide, which is, even if it is the case,
    a hard problem by itself since the method we are using make it possible
    to find only an upper bound of $\Omega$ rather than its precise value.
    Even more important is that the spectrum of our problem, actually posed
    in the whole Jordan space-time, may in principle be different from those
    of the ``partial'' problems formulated separately in its two regions. So
    we invoke a coordinate that covers the whole Jordan manifold, and we are
    dealing with a single operator. The circumstance that the whole metric
    (\ref{dsEy}) changes its sign at \Str\ is inessential since this metric
    only plays an auxiliary role.

    Let us bring the differential equation (\ref{st1q}) to a self-adjoint
    Sturm-Liouville form:
\beq
    -(p\ \delta \beta')' + q\ \delta\beta = -E r\,\delta\beta,
                                \label{Sturm}
\eeq
    where
\bearr
      p = \frac{y(1+y)(1 + y_1 + 2 y_1 y)^2 (1 + 2y)^2}
                {(1 + y_1 + 6y_1 y + 12 y_1y^2 + 8y_1 y^3)^2},
\nnn
      q = \fract{1}{4}[\fract 1 8 + (y + \half)^3 y_1)^3]^{-1}
       (y + \half) [\half + (y+\half)y_1]
\nnn\ \
     \times [(y+\half)^3
                y_1^2 + (2y^3 + y^4 + \half y
            + \fract 3 2 y^2) y_1 - \fract 1 8 - \fract 1 4 y],
\nnn
      r= \frac{y (1 + y) [\half  + (y + \half ) y_1]^6}
        {[\fract 18 + (y + \half)^3 y_1]^2 (y + \half)^2 y_1^4},
\nnn
      E = - \Omega^2 h^2.
\earn

    It can be shown that when the \wh\ charge tends to zero ($y_1 \to 1$),
    this equation reduces to that for an uncharged \wh\ \cite{worm3}. In
    other words, using the linear operator perturbation theory, one can show
    that, for small charges, our boundary-value problem has solutions with
    values of $E$ close to the ones for neutral \whs.

    Let us formulate the boundary conditions. Since our equation covers
    the whole manifold, we have two conditions at two spatial asymptotics:
    $\delta\beta \to 0$ at both. At the transition sphere \Str, the solution
    has the approximate form
\beq
     \delta\beta \approx c_1 + c_2\ln (y-1).
\eeq

    It is now necessary to find out whether or not there are solutions with
    $E < 0$ and $c_2 = 0$, vanishing at infinity.

    \eq (\ref{Sturm}) is brought to a \Srd\ form by the substitution
\bear                                     	\label{transform}
        x \eql \frac{1}{m} \int s^2 d y,
\nn
     \delta\beta  \eql  \frac{z}{s} \exp \left(-\Half \int F d u\right),
\ear
    The result is
\beq                                         	\label{schrod_q}
        d^2 z(x)/dx^2 + [E - V(x)] z(x) =0.
\eeq
    with a potential $V$ having the asymptotics
\bearr   	                             \label{pot1}
     V(x) \approx \frac{2}{y_1\ x^3} \qquad
            (x\to \pm \infty, \ \ \mbox{flat asymptotics}),
\nnn
     V(x) \approx \widetilde{V}(x) =
         -\frac{1}{(4x^2)}-\frac{11\ y_1^2(y_1-1)}{4(1+y_1)^3\ x}
\nnn  \inch
                (x\to 0, \quad \Str).
\ear
    The complete form of the potential is very cumbersome and, on the other
    hand, unnecessary for the further study.

    At the transition sphere, the asymptotic form of the solution is
\beq
    z \simeq  c_1 \sqrt{x}(1+b\,x)+c_2\ln{x} \sqrt{x} (1+b\, x),
\eeq
    where
\[
    b = -\frac {11}{4} \frac{y_1^2(y_1-1)}{(y_1+1)^3}.
\]

    The further reasoning is carried out similarly to that of the previous
    work \cite{worm3}. Namely, we study the properties of the \Srd\ operator
    corresponding to \eq (\ref{schrod_q}) for small values of the charge
    $q$, i.e., for small $b$. We prove that this operator is self-adjoint,
    and its continuous spectrum covers the whole non-negative semiaxis.
    Thus an instability, if any, corresponds to discrete ``energy'' levels.
    A further analysis, which is rather technically complicated and includes
    the use of linear operator perturbation theory \cite{RS, Kato}, leads to
    the conclusion that there exists at least one negative ``energy level''
    $E$.

    We have thus shown that, for small values of the charge parameter $b$,
    our boundary-value problem has solutions describing exponentially
    growing perturbations with increment values close to those for uncharged
    \whs.

    This instability conclusion is applicable to any STT (\ref{act})
    admitting \wh\ solutions due to conformal continuation.

    Numerical calculations have shown that the perturbation increment
    diminishes as the charge grows. For large values of the charge (compared
    to the \wh\ radius in proper units), the numerical methods used become
    unreliable, but the available results make us suppose that at larger
    charges the \whs\ stabilize with respect to \sph\ perturbations.

    To conclude, we give an example of a numerical estimate. In the case
    of conformal coupling, the throat radius is approximately equal to $2h$.
    The characteristic time of decay, $\tau =1/\Omega$, is proportional to
    $h$ (which has the dimension of length):
\beq                            \label{tau}
               \tau \simeq h /\sqrt{0.048} \simeq 5 h.
\eeq
    For a wormhole radius of the order of a typical stellar size $\sim
    10^6$ km, the time $\tau$ is a few seconds, slightly greater than the
    time needed for a light signal to cover the stellar diameter.

    Similar estimates can be obtained for other STT characterized by
    different $f(\phi)$.

\end{document}